\documentstyle[12pt,epsfig,here]{article}
\begin{document}
\begin{sloppypar}
\begin{center}

{\bf PRELIMINARY RESULTS ON A SEARCH FOR NEUTRINOS}
{\bf FROM THE CENTER OF THE EARTH}
{\bf WITH THE BAIKAL UNDERWATER TELESCOPE}

\vspace{4mm}

{\footnotesize

L.B.Bezrukov, B.A.Borisovets, I.A.Danilchenko, J.-A.M.Djilkibaev, 
G.V.Domogatsky, A.A.Doroshenko, A.A.Garus, A.M.Klabukov, S.I.Klimushin,
B.K.Lubsandorzhiev, A.I.Panfilov, D.P.Petukhov, P.G.Pokhil, I.A.Sokalski \\
{\em Institute for Nuclear Research,  Russian Academy of Sciences (Moscow, 
Russia) \\}
\smallskip
N.M.Budnev, A.G.Chensky, V.I.Dobrynin, O.N.Gaponenko, O.A.Gress,
T.A.Gress, A.P.Koshechkin, R.R.Mirgazov, A.V.Moroz, S.A.Nikiforov,
Yu.V.Parfenov, A.A.Pavlov, K.A.Pocheikin, P.A.Pokolev, V.Yu.Rubzov,
S.I.Sinegovsky, B.A.Tarashansky \\
{\em Irkutsk State University (Irkutsk, Russia) \\}
\smallskip 
S.B.Ignat$'$ev, L.A.Kuzmichov, N.I.Moseiko, E.A.Osipova \\
{\em Moscow State University  (Moscow, Russia) \\}
\smallskip 
S.V.Fialkovsky, V.F.Kulepov, M.B.Milenin \\
{\em Nizhni Novgorod State Technical University  (Nizhni Novgorod, Russia) \\}
\smallskip
                                  M.I.Rozanov \\
{\em   St.Petersburg State Marine Technical University (St.Petersburg, 
Russia) \\}
\smallskip 
                                  A.I.Klimov \\
{\em                     Kurchatov Institute (Moscow, Russia) \\}
\smallskip 
                               I.A.Belolaptikov \\
{\em             Joint Institute for Nuclear Research  (Dubna, Russia) \\}
\smallskip 
       H.Heukenkamp$^*$, A.Karle, T.Mikolajski, Ch.Spiering,
               O.Streicher, T.Thon, Ch.Wiebusch, R.Wischnewski \\
{\em           DESY Institute for High Energy Physics (Zeuthen, 
Germany),  $^*$now at University of Wisconsin}
\vspace{3mm}
}
{\small
{\sf presented by J.Djilkibaev}

{\sf to the 2nd Workshop on 
"Dark Side of the Universe: Experimental efforts and theoretical framework",
Rome, Italy, November 13-14 1995}
}
\end{center}
\vspace{-4mm}
\begin{abstract}
{\footnotesize 

The deep underwater Cherenkov neutrino telescope NT-200 is currently under
construction at lake Baikal. Its first stage  \mbox{NT-36} consisting of 36 
optical modules has operated over 2 years since April 1993 till March 1995.
Here we present a method to search for nearly vertical upward going muons 
from neutralino annihilation in the center of the Earth. We present preliminary
results obtained from experimental data taken with the \mbox{NT-36} array
in 1994.}
\end{abstract}
\vspace{3mm}
{\bf 1. Introduction}

\vspace{3mm}
An attractive way to search  for cold dark matter is the detection of 
high-energy neutrinos produced  by neutralino annihilation in the Earth and 
in the Sun. The Baksan and the Kamiokande collaboration have  presented 
stringent limits on the up-going muon flux initiated by neutralinos in the 
Earth$^{1,2}$. Further progress is possible with underwater (BAIKAL, DUMAND, 
NESTOR) and under ice \mbox{(AMANDA)} experiments which will have effective 
areas of 2-10 thousand square meters. The first large deep underwater detector
for muons and neutrinos, \mbox{NT-200}, is currently under construction in the
lake Baikal. The first stage of the detector  consisting of 36 optical  modules
\mbox{(NT-36)} successfully has operated over 2 years since 1993 till 1995.

The first attempt for an indirect search for neutralinos with
an underwater experiment has been performed on the base of experimental data 
taken during 1994 with the detector \mbox{NT-36.} 
\vspace{5mm}
\begin{par}
\noindent
{\bf 2. Detector}
\vspace{2mm}
\end{par}

The Baikal Neutrino Telescope$^{3}$ is being deployed in the Siberian Lake 
Baikal, about 3.6 km from the shore at the depth of 1.1 km. It will consist 
of 192 optical modules (see Fig.1). The 7 arms of the umbrella-like frame 
carrying the detector, each 21.5 m in length, are at the height of 250 m above
the bottom of the lake. The optical modules are grouped in pairs along the 
strings, directed alternatively upward and downward. The distance between 
pairs looking face to face is 7.5 m, while pairs arranged back to back are 5 m
apart. The pulses from two PMTs of a pair after \mbox{0.3 {\it p.e.}} 
discrimination are fed to a coincidence circuit with \mbox{15 ns} time window.
A PMT pair defines a {\it channel} with its output denoted as {\it local 
trigger} (or simply {\it hit}). A {\it muon-trigger} is formed by the 
requirement of \mbox{$\geq 3$ hits} within a time window of \mbox{500 ns}. For
such events, amplitude and time of all hit channels are digitized and sent 
to shore.

\begin{figure}[H]
\centering
\mbox{\epsfig{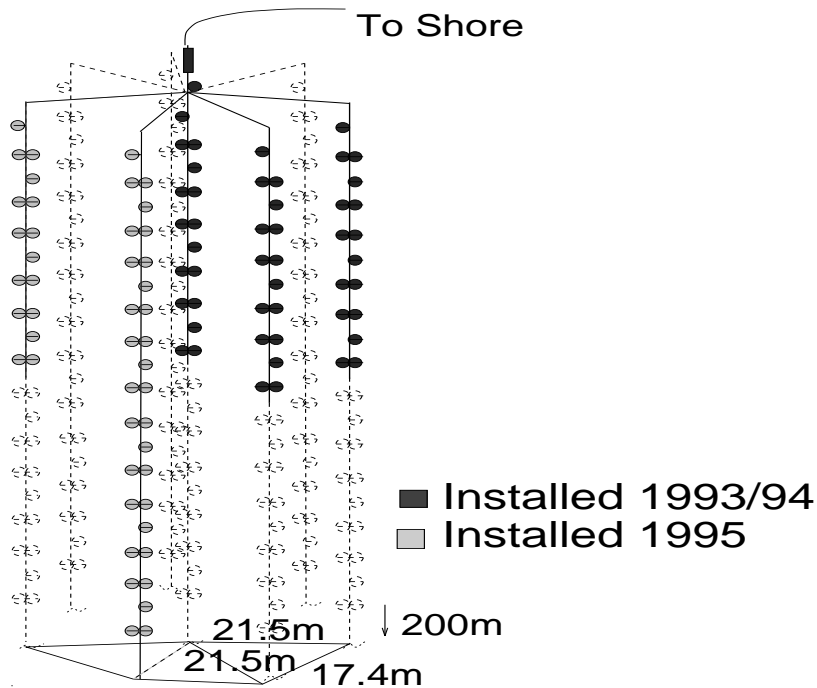}}
%%%\hspace{2.5cm}
%%%\parbox[b]{4cm}{\caption [1]
\parbox{15cm}{\caption [1]
         {\footnotesize Schematical view of the planned NT-200 detector. \mbox{NT-36} components
             operating since April 1993 are in black, additional modules
             deployed in March 1995 in grey.}                    
               }
\end{figure}

In  April  1993, NT-36 started data taking and operated till March 1995. 
There have been 6 PMT pairs along each of 3 strings of \mbox{NT-36.} 
The numbering is, from top to bottom: 1(up-looking), 2(down-looking), 
3(up), 4(down), 5(up), 6(down) with PMT orientations which are 
given in the parenthesis for 1994 \mbox{NT-36} modification. 
\vspace{5mm}
\begin{par}
\noindent
{\bf 3. Method}
\vspace{2mm}
\end{par}

Our search for possible high-energy neutrino events resulting from dark matter
annihilation in the center of the Earth is based on the analysis of the 
experimental data with respect to upward-going
muons within a cone of about 15 degree half-aperture around the opposite 
zenith.

\vspace{-6.5mm}
\begin{figure}[H]
\centering
\mbox{\epsfig{file=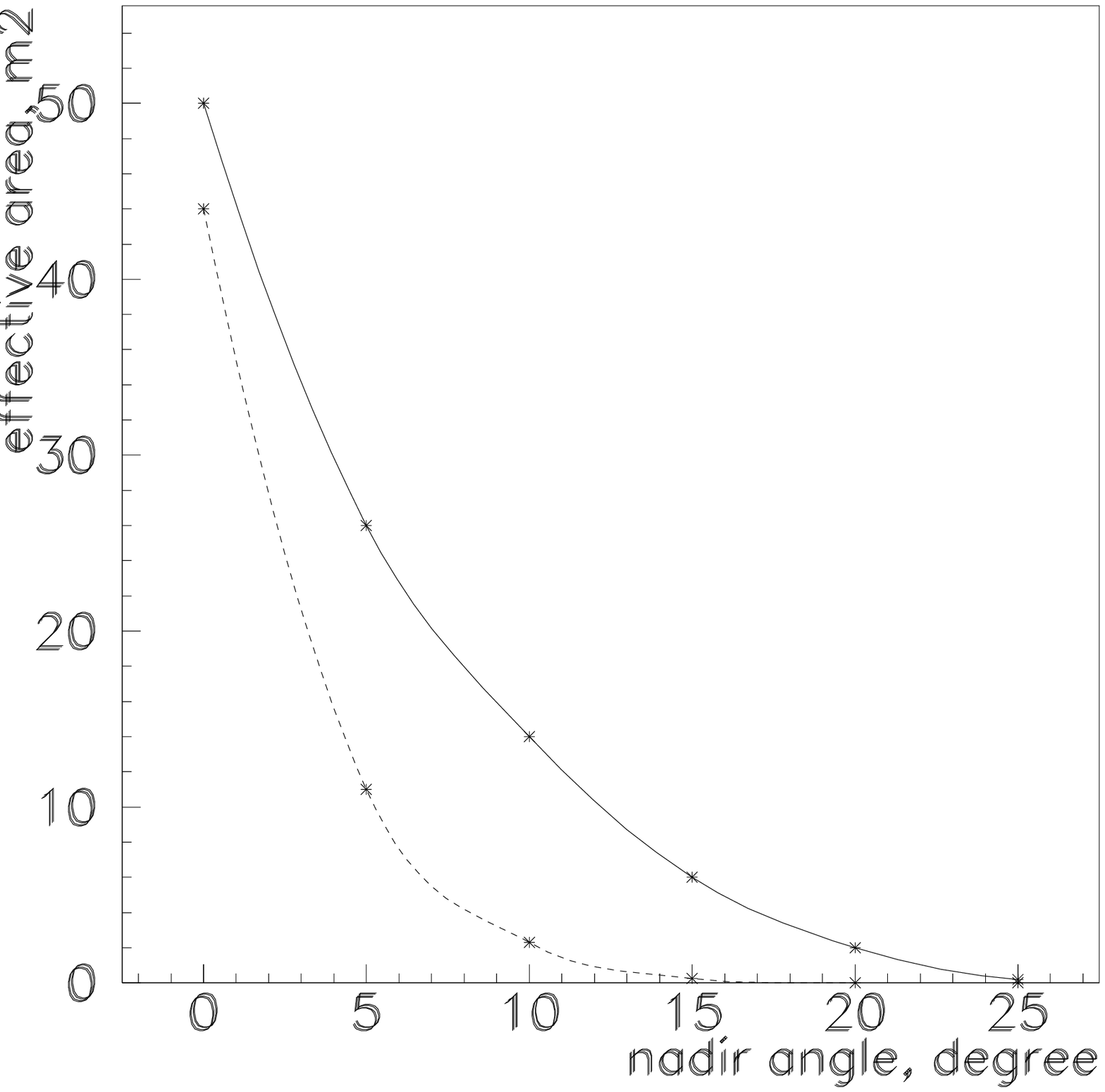,width=7.2cm,height=8cm}}
\mbox{\epsfig{file=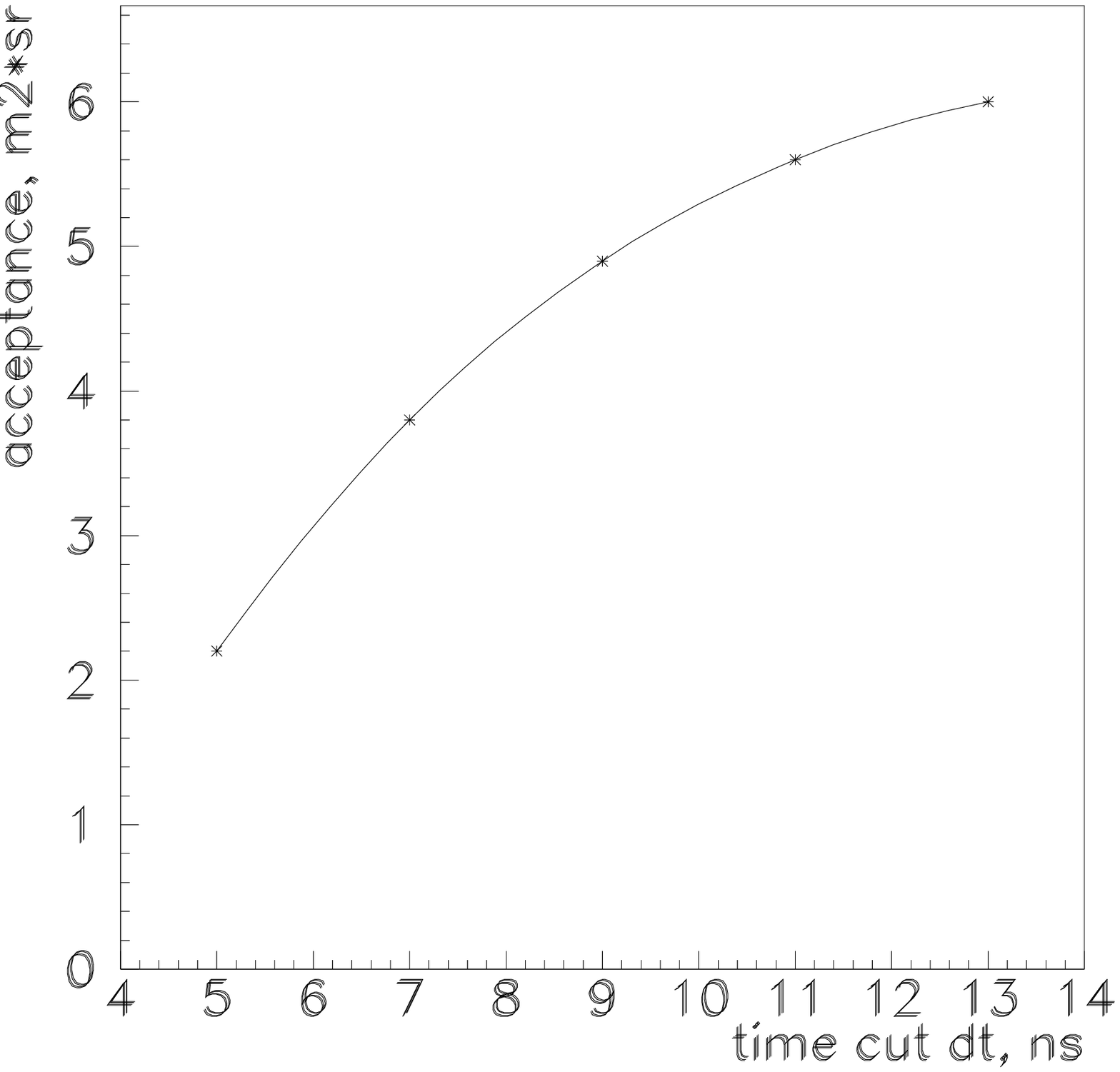,width=7.2cm,height=8cm}}
     \parbox{7cm}{
          \caption [2]
        {\footnotesize The single string effective area vs. muons zenith angle.
Solid and dashed lines correspond to $dt$ = 13 and 5 ns, respectively. 
               }
}
\hspace{.5cm}
     \parbox{7cm}{
          \caption [3]
    {\footnotesize The single string acceptance vs. time cut values $dt$.
}
          }
\end{figure}

In contrast with  our standard reconstruction strategy, which supposes 
\mbox{$\geq 6$} hits at \mbox{$\geq 3$ strings} (necessary for full spatial 
reconstruction$^{4}$), we did not perform a reconstruction at all, 
but applied cuts, which effectively reject all events with the exception
of nearly vertically moving upward muons triggering exclusively
channels along the string. In our case this method works well since the tracks 
of the objects searched for have nearly the same vertical orientation as the 
strings. The following off-line trigger conditions were used:

\begin{description}
 \item[\#1]
three down-looking and at least one up-looking channels at one string 
exclusively must be hit; 

 \item[\#2]
time differences of any two hit channels $i$ and $j$ must obey the inequality
             \[abs((t_{i}-t_{j})-(T_{i}-T_{j}))<dt\]
 - where $t_{i}(t_{j})$ is an experimental time of channel $i(j)$,
             $T_{i}(T_{j})$ is the "theoretical" time expected for minimal 
ionizing, up-going vertical
      muons and $dt$ is a time cut \mbox{($dt=(5,7, \ldots ,13, \ldots )$ns)};
 \item[\#3]
      the minimum value of amplitude asymmetries for all pairs of alternatively
        directed (up/down) hit channels must obey the 
inequality $dA_{ij}(down-up)|_{min}> 0.3$, where
       \[dA_{ij}(down-up)= (A_{i}(down)-A_{j}(up))/(A_{i}(down)+A_{j}(up)),\] 
$A_{i}(down)(A_{j}(up))$ 
- amplitude of channel $i(j)$ looking downward(upward);
 \item[\#4]
     the maximum value of amplitude asymmetries for all pairs of down-looking 
  hit channels must obey the inequality $dA_{ij}(down-down)|_{max}< 0.7$ where
\[dA_{ij}(down-down)= abs(A_{i}(down)-A_{j}(down))/(A_{i}(down)+A_{j}(down)).\]
 \end{description}

The effective area and acceptance of NT-36 for off-line 
trigger conditions \#1--\#4 obtained from MC   
        simulations are presented in Fig.2 and Fig.3.

\vspace{5mm}
\begin{par}
\noindent
{\bf 4. Results}
\vspace{2mm}
\end{par}

Our analysis is based on the data taken during the time period April 8 to 
       \mbox{November 9}, 1994.
        This corresponds to 150 days of detector live time. Upward-going muon
        candidates were selected from a total of 
\mbox{$7.72\cdot 10^{7}$} events recorded during this period
        by the muon-trigger \mbox{$\geq 3$.} Eight events fulfill the 
earlier defined off-line trigger conditions
\#1--\#3 with $dt=13$ns. Seven events of this sample have a value of maximum
 asymmetry for down-looking channels \mbox{$dA_{ij}(down-down)|_{max}>0.8$} 
and have been classified as 
showers, generated by downward going atmospheric muons. 
Only 1 event fulfills all off-line trigger conditions
        \#1--\#4 with $dt=13$ns. It is 6 June 1994 event. 
Experimental and "theoretically" expected (in parenthesis) values of time 
differences for hit channels in this event are presented in Tab. 1. 
The values of amplitude asymmetries for up-down and down-down looking 
channel combinations are presented in Tab. 2. 

{\footnotesize 
\begin{tabbing}
jjjjjjjjjjjjjjjjjjjjjjjjjjjjjjjjjjjjjjjjjjjjjjjjjjjjjjj \= 
jjjjjjjjjjjjjjjjjjjj
jjjjjjj \= jjjjjjjjjjjjjjjjjjjjjjjjjjjjjjj \kill 
        Table 1. Experimental and ``theoretically'' \\  
expected (in parenthesis) values of time   \>   \> Table 2. 
The values of amplitude \\
        differences between hit channels $i,j$. \>   \> 
asymmetry for hit channels $i,j$.\\
\end{tabbing}
}
\vspace{-0.7cm}
\begin{par}
\noindent
	\begin{tabular}{|c|c|c|c|c|c|c|c|c|} \cline{1-4} \cline{6-9}
	$i / j$   & ch.4   & ch.5 & ch.6   & \ \ \ \ \ \ \ \ \ \ \ \ \ \ \  
& $i / j$   & ch.4   & ch.5 & ch.6   \\ 
                & (down) & (up) & (down) &   &       
& (down) & (up) & (down) \\ \cline{1-4} \cline{6-9}
ch.2    & 44ns   & 64ns & 87ns  &   & ch.2  & 0.05  & 0.64  & 0.26  
\\               
	(down)  & (42ns) & (67ns) & (84ns)  &   & (down)  
&       &       &       \\ \cline{1-4} \cline{6-9}
	ch.4    &  ---   & 21ns & 43ns &   & ch.4  
& ---   & 0.66  & 0.30  \\            
	(down)  &        & (25ns) & (42ns)  &   & (down)  
&       &       &       \\  \cline{1-4} \cline{6-9}
	ch.5    & ---    & ---    &  22ns &   & ch.5    
& ---   & ---   & 0.45  \\ 
	(up)    &        &        & (17ns)  &   &  (up)  
&       &       &       \\ \cline{1-4}  \cline{6-9}
	\end{tabular} 
\end{par}
\medskip

The time pattern of such event might be generated by a shower
below the detector or by a nearly horizontal muon 
(being rare events themselves). 
However, it is difficult to imagine that something else 
but an upward going muon could 
generate the observed amplitude pattern in combination with the time one.
Thorough MC calculations in order to ascertain the background due to
fake events are underway. 
Some data corrections are still necessary concerning sedimentation, time 
stability of PMTs sensitivity etc.
But our preliminary estimations yield the signal-to-fake 
ratio of the order of unity 
for this kind of events. Therefore, we consider the event as 
the first promising 
neutrino candidate. 

The expected number for events generated by upward going muons from 
atmospheric neutrinos was calculated as 0.4 for runtime \mbox{0.41 year.}
Then, regarding the only neutrino candidate as an atmospheric 
neutrino event, an \mbox{90 $\%$ CL} upper limit of 
\mbox{$2.5\cdot 10^{-13}$ (muons/cm$^{2}$/sec)} 
in a cone with 15 degree half-aperture 
around the opposite zenith is obtained for 
upward going muons generated by neutrinos 
due to neutralino annihilation in the center of the Earth.
This limit corresponds to muons with energies greater than threshold
energy \mbox{$E_{th}\approx$ 6 GeV,} defined by the 30m  string length.  

\vspace{5mm}
\begin{par}
\noindent
{\bf 5. Conclusions}
\vspace{2mm}
\end{par}

We have presented the preliminary analysis of experimental data taken with the
underwater detector \mbox{NT-36} in order to study the capability of 
the indirect search
for dark matter with Baikal experiment. 
The first promising candidate for an upward going muon was identified.
	An upper limit for the up-going muon flux has been obtained.
This is still an order of magnitude higher than the limits obtained by
Baksan and Kamiokande. 
The effective area of \mbox{NT-36} for nearly vertical up-going muons
fulfilling our off-line trigger conditions is 
\mbox{$S_{eff}$ = 50 m$^{2}$/string.}  
A rough estimation of the effective area
of the full-scale Baikal neutrino telescope 
\mbox{NT-200} (comprising eight twice longer strings)
        for detection of upward going muons 
%in a narrow cone around the center of Earth 
gives the value \mbox{$S_{eff} \approx$ 400-800 m$^{2}$.}

\vspace{5mm}
\begin{par}
\noindent
{\bf Acknowledgements}
\vspace{2mm}
\end{par}

Three of us (J.D., I.S., G.D.) are grateful to S.Mikheyev 
for helpful discussions. 
We also acknowledge L.Pavlova for technical assistance.

\vspace{3mm}
{\small
{\bf References}
\begin{enumerate}
\vspace{-1mm}
\item M.Mori et al., {\em Phys. Rev.} {\bf D48} (1993) 5505.
\vspace{-3mm}
\item M.Boliev et al., {\em Proc. Int. Workshop on Theor. and Phenom. Aspects 
of Underground Phys.}
(Toledo, 1995), to be published.
\vspace{-3mm} 
\item I.A.Belolaptikov et al., {\em Proc. Third Int. Workshop on Neutrino 
Telescops}
(Venice, 1991) 365; I.A.Belolaptikov et al., {\em Nucl. Phys.} {\bf B19} 
(1991) 375; I.A.Belolaptikov et al., {\em Proc. 23rd ICRC} vol.4 
(Calgary 1993) 573; I.A.Sokalsky and Ch.Spiering (eds.), {\em The 
Baikal Neutrino Telescope NT-200, BAIKAL 92-03}
(1992); I.A.Belolaptikov et al., {\em Proc. 24rd ICRC} vol.1 (Rome 1995) 742.
\vspace{-3mm}
\item I.A.Belolaptikov et al., {\em Nucl. Phys.} {\bf B35} (1994) 301.
\end{enumerate}
}

\end{sloppypar}
\end{document}